\documentclass{epl}
\usepackage{graphics}
\usepackage{epsfig}
\usepackage{amssymb}
\usepackage{amsmath}
\usepackage{bbm}

\title{Entanglement of $2 \times K$ quantum systems}
\author{Artur {\L}ozi{\'n}ski\inst{1,2} \and Andreas Buchleitner\inst{2} 
\and Karol {\.Z}yczkowski\inst{1,3} \and Thomas Wellens\inst{2}}
\institute{
  \inst{1} Instytut Fizyki im. Smoluchowskiego,  Uniwersytet
Jagiello{\'n}ski,  ul. Reymonta 4, PL-30-059 Krak{\'o}w\\
  \inst{2} Max-Planck-Institut f\"ur Physik komplexer Systeme, N\"othnitzer
                                Stra{\ss}e 38, D-01187 Dresden\\
  \inst{3} Centrum Fizyki Teoretycznej, Polska
  Akademia Nauk, Al. Lotnik{\'o}w 32/46, PL-02-668 Warszawa
}

\begin{document}

\maketitle

\begin{abstract}
We derive an analytical expression for the lower bound of the
concurrence of mixed quantum states of composite $2\times K$ systems. In
contrast to other, implicitly defined entanglement measures, the
numerical evaluation of our bound is straightforward. We explicitly
evaluate its tightness for general mixed states of $2\times 3$ systems,
and identify a large class of states where our expression gives the
exact value of the concurrence.
\end{abstract}

\section{Introduction}

Entanglement is a fundamental concept in the theory of  quantum
information, and the essential resource for (potential) modern
applications of quantum mechanics \cite{IQCI,NC00}. As the cause of
nonclassical correlations between measurement results on the different
constituents of multipartite quantum systems it is the key ingredient
for  quantum cryptography, teleportation, and quantum computing.
However, the quantitative characterization of entanglement remains a
largely open problem, due to the ever more intricate topology and
rapidly increasing dimension of the space of admissible quantum states
with the number of components of composite quantum systems.  By now,
only the entanglement of pure bipartite states is well understood and
unambiguously quantified \cite{Be96,BB96,Ni99,Wo98}. For mixed states --
which we generically encounter in nature -- a large variety of
entanglement measures \footnote{The required properties of an
entanglement measure are: monotonously decreasing evolution under local
operations with classical communication; convexity on state space;
vanishing values for separable states alone \cite{Be96,Ve97}.} was
proposed (see, e.g., \cite{Ho01}). Almost none of them, however, can be
computed efficiently, the only exceptions (to our knowledge) being the
{\em negativity} \cite{ZH98,Vi02}, and the {\em entanglement of
formation} \cite{Be96} of two qubits \cite{Wo98}.  For instance, to
calculate the entanglement of formation in higher dimensional spaces,
one needs to perform a complicated optimization procedure over the huge
space of all possible decompositions of the analyzed mixed state $\rho$
into pure states, such as to minimize the average pure state
entanglement needed to represent $\rho$.  By construction, such a
procedure can only provide an upper bound for the degree of entanglement
of $\rho$ \cite{Zy99,Ad00,Tu01}.  What is needed in addition is a lower
bound of mixed state entanglement, which we shall provide for $2\times
K$ systems in the present contribution. To do so, we will use the {\em
concurrence}, originally introduced by Wootters \cite{Wo98} for $2\times
2$ states, and generalized for any bipartite pure states by Rungta {\em
et al.} \cite{RB01}. (Alternative generalizations were proposed in
\cite{Ul00,BD01}.) We shall generalize its definition for general
$2\times K$ mixed states and derive an analytical lower bound which also
implies a lower bound for the entanglement of formation.  For some class
of $2\times 3$ mixed states we can establish equality between our lower
bound and the exact value of the concurrence, and numerical calculations
assess the tightness of our bound for general $2\times 3$ mixed states.

\section{Theory}
Let us start with the definition of the concurrence of a pure state on a
$N\times K$ dimensional Hilbert space ${\cal H} = {\cal H}_N \otimes
{\cal H}_K$ \cite{RB01}.  We first need the {\em flip operator} $\cal F$
acting on an arbitrary Hermitian operator $A$ on ${\cal H}$,
\begin{equation}
{\cal F}(A):= A+({\rm tr} A){\mathbb I}- ({\rm tr}_N A) \otimes
{\mathbb I}_{K} -{\mathbb I}_{N}\otimes ({\rm tr}_K A)\, , 
\label{flip} 
\end{equation}
with ${\mathbb I}_K$ and ${\mathbb I}_N$ the identity on ${\cal H}_K$
and ${\cal H}_N$, respectively, and ${\rm tr}_{K}$ (${\rm tr}_N$) the
partial trace over ${\cal H}_{K}$ (${\cal H}_{N}$).  $\cal F$ commutes
with all unitary operators and preserves positivity.  Moreover, the
expectation value $\langle \psi|{\cal F}(\rho_{\psi}) |\psi\rangle$,
where $\rho_{\psi}~=~|\psi\rangle \langle \psi|$, is non-negative for
all pure states and equals zero if and only if $|\psi\rangle \in {\cal
H}^N\otimes {\cal H}^K$ is a product state.  This allows to define the
concurrence of any arbitrary bipartite pure state 
as \cite{RB01} 
 \begin{equation}
 C(|\psi\rangle) = 
 \sqrt{\langle \psi|{\cal F}(\rho_{\psi})|\psi\rangle}\, ,
\label{concgen} 
\end{equation}
wherefrom 
\begin{equation}
C(|\psi\rangle) = \sqrt{2\bigl[|\langle \psi | \psi \rangle |^2-{\rm
tr(\rho_N^2)} \bigr]}
\label{concge2} 
\end{equation}
is easily deduced, with $\rho_N = {\rm tr}_K |\psi\rangle \langle \psi|$
the reduced density operator of dimension $N$. Observe that $C$ is
linear in the norm of $|\psi\rangle$. For a normalized state, $\langle
\psi | \psi \rangle =1$, it interpolates monotonously between  zero for
product states, and $\sqrt{2(N-1)/N}$ for maximally entangled states,
where $N\leq K$ is assumed (without loss of generality).

We now need the generalization of this definition for mixed states. As a
consequence of the generalized construction outlined in \cite{Ul00}, any
entanglement measure on the pure states can be ported to the mixed
states through minimization of its average value on all possible
decompositions into pure states.  For instance, the {\em entanglement of
formation} $E_F(\rho )$ of a mixed state is defined \cite{Be96} as the
infimum over all possible pure state decompositions $\sum_{l=1}^L p_l
|\psi_l \rangle \langle \psi_l |$, 
\begin{equation}
E_F(\rho):= {\rm inf}_{\cal E} \sum_{l=1}^L p_l E(|\psi_l\rangle ),
\label{eof} 
\end{equation}
with $E(|\psi \rangle) = -{\rm tr} (\rho_N \ln \rho_N )$ the {\em
entropy of entanglement}.  The length of the ensemble $L$ is arbitrary,
but need not exceed $(NK)^2$ \cite{Ul00}.  In close analogy, we define
the mixed state concurrence as
\begin{equation}
C(\rho):= {\rm inf}_{\cal E} \sum_{l=1}^L p_l C(|\psi_l\rangle ).
\label{concge3} 
\end{equation}

Note that, in the simplest case of $2\times 2$ systems, the infima in
(\ref{eof}) and (\ref{concge3}) can be realized simultaneously, by the
same optimal decomposition, and the definition (\ref{concge3}) thus
reduces to the original one \cite{Wo98}.  However, in the general case
of $N\times K$ systems the minimum average entropy of entanglement need
not be achieved with the same (optimal) decomposition as the minimum
average concurrence~\cite{Ul00}. 

To derive an lower bound of $C(\rho )$ from below, we now specialize to the
case $N=2$.  If a $2\times K$ state is pure, its concurrence may be
expressed in terms of unnormalized states $P^{(ij)}|\psi\rangle$, which
live on a $2\times 2$ dimensional Hilbert space.  Let $\{ |k\rangle
\}_{k=1,\dots,K}$ be an arbitrary basis of ${\cal H}_K$, and 
\begin{equation}
P^{(ij)}~=
{\mathbb I}_2 \otimes  (|i\rangle\langle i|+|j\rangle\langle j|)\; ,\ i,j\in\{
1,\ldots ,K\}\; ,
\end{equation}
denote the projections onto $2\times 2$ dimensional subspaces.  We then
obtain the squared concurrence of a $2\times K$ pure state directly from
the definition (\ref{concgen}):
\begin{equation}
C^2(|\psi\rangle)=\sum_{i=1}^K \sum_{j=i+1}^K
C^2\left(P^{(ij)}|\psi\rangle\right).
\label{2kto22pure}
\end{equation}
Given this expression, we can  derive the desired lower bound.
According to 
(\ref{concge3}),
\begin{equation}
C(\rho)~=~\sum_{l=1}^L p_l C(|\psi_l\rangle)
\end{equation}
for the {\em optimal decomposition} $\{ |\psi_l\rangle \}$.  Insertion
of (\ref{2kto22pure}) yields
\begin{equation}
C(\rho)~=~\sum_{l=1}^L p_l \sqrt{\sum_{i>j}
C^2\left(P^{(ij)}|\psi_l\rangle\right)}~=
~\sum_{l=1}^L p_l \sqrt{\sum_{i>j}r_{ij}
  C^2\left(\frac{P^{(ij)}|\psi_l\rangle}{r_{ij}^{1/4}}\right)}\; ,
\label{concave1}
\end{equation}
where we introduced arbitrary coefficients $r_{ij}\geq 0$, with
$\sum_{i>j} r_{ij}=1$. Next, we use the concavity of the square root and
obtain:
\begin{equation}
C(\rho)~\geq~\sum_{l=1}^L p_l \sum_{i>j} r_{ij}
\sqrt{C^2\left(\frac{P^{(ij)}|\psi_l\rangle}{r_{ij}^{1/4}}\right)}~=
\sum_{i>j} \sqrt{r_{ij}}\sum_{l=1}^L p_l~C\left(P^{(ij)}|\psi_l\rangle\right).
\label{concave2}
\end{equation}
The sum over $l$ on the right hand side can be interpreted as the
average concurrence $\langle C\rangle$ of the $2\times 2$ density
matrices $\rho^{(ij)}$ obtained by projections
\begin{equation}
\rho^{(ij)}~=~P^{(ij)}\rho P^{(ij)}~= ~\sum_{l=1}^L p_l
P^{(ij)}|\psi_l\rangle\langle\psi_l|P^{(ij)}.
\label{decompij}
\end{equation}
Due to the definition (\ref{concgen}), $\langle C\rangle$ cannot be
smaller than the projected concurrence $C(\rho^{(ij)})$, which is known
analytically as the concurrence of a $2\times 2$ mixed state.  Hence, 
\begin{equation}\label{ineqC1}
C(\rho)\geq \sum_{i=1}^K \sum_{j=i+1}^K
\sqrt{r_{ij}}~C\left(\rho^{(ij)}\right).
\end{equation} 
Since the $r_{ij}$ were not fixed in (\ref{concave1}), we still can
choose these coefficients $r_{ij}$ such as to maximize the sum on the
right hand side of 
(\ref{ineqC1}),
\begin{equation}
r_{ij}~=~\frac{C^2\left(\rho^{(ij)}\right)}
{\sum_{m>n}C^2\left(\rho^{(mn)}\right)}\label{weights}\, ,
\end{equation}
and arrive at
\begin{equation}
C(\rho)\geq \sqrt{\sum_{i>j}~C^2\left(\rho^{(ij)}\right)}.
\label{lowerbound}
\end{equation} 
This lower bound still depends on the choice of the basis $\{ |k\rangle
\}_{k=1,\dots ,K}$, of the $K$ dimensional subsystem.  To find the
tightest lower bound, we have to maximize (\ref{lowerbound}) over all
orthogonal basis sets, tantamount to finding the unitary $K\times K$
operator $U$ which transforms to the optimal basis $\{U|k\rangle
\}_{k=1,\dots ,K}$.  Note that such a search is much faster than finding
the optimal decomposition of the mixed state via a suitable unitary
transformation of dimension $4K^2$.

Finally, since the entropy of entanglement of a pure state on $2\times
K$ systems is a convex function of its concurrence \cite{Wo98},
\begin{equation}
E(|\psi\rangle ) = h \left
(\frac{1+\sqrt{1-C(|\psi\rangle)^2}}{2}\right )\; , \quad 
h(x) = -x\log_2 x - (1-x)\log_2 (1-x)\; ,
\end{equation}
the inequality (\ref{lowerbound}) immediately provides a lower bound for
the entanglement of formation defined in 
(\ref{eof}):
\begin{equation}
E_F(\rho ) \geq 
h\left [ \frac{1}{2}\left (1+\sqrt{1-\sum_{i>j}C^2(\rho^{(ij)})}\right )
\right ]\,    .
\end{equation}
As a matter of fact, the same bound was equally proposed in \cite{Ch02}.
However, the proof in \cite{Ch02} is invalid, since it relies on the
incorrect assumption that the entanglement of formation is a convex
function of the {\em squared} concurrence $C^2$.
\begin{figure}
\begin{center}
\leavevmode
\epsfxsize14cm
\epsfbox{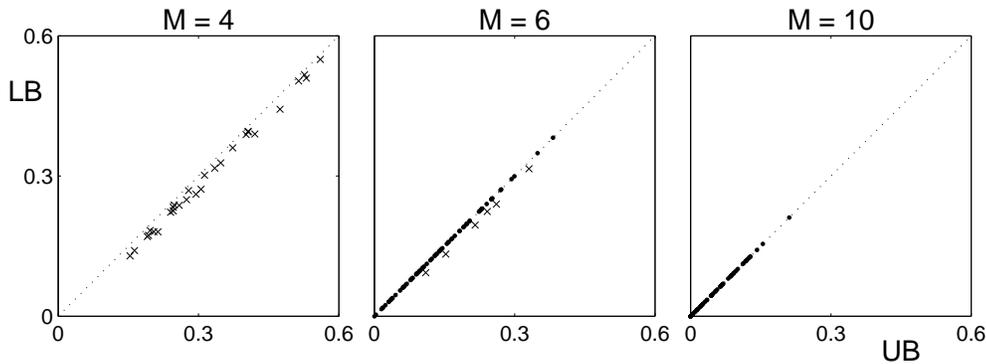}
\end{center}
\caption{Lower bound ({\it LB}), eq.~(\ref{lowerbound}), versus upper
bound ({\it UB}), eq.~(\ref{concge3}), of the concurrence for ensembles
of randomly chosen  $2\times 3$ mixed states, generated by a partial
trace over pure states in a $6M$ dimensional state space, with
$M=4,6,10$. The deviation from the diagonal indicates the tightness of
both bounds, which is obviously very good over the entire range of the
three random ensembles.  Dots ($\bullet$) identify mixed states for
which the lower bound provides the exact value.}
\label{ublbfig}
\end{figure}

Given the above estimate, we shall now assess its tightness for the
special case $K=3$. In this case, (\ref{lowerbound}) allows to
distinguish between entangled and separable states, since any entangled
$2\times 3$ state also has entangled projections onto $2\times 2$
subspaces.  (This is not the case for $K>3$, due to the existence of
bound entangled states with positive partial transpose \cite{Hos98}.)
Our random mixed states were generated by partial trace ${\rm
tr}_M(|\chi \rangle \langle \chi |)$ over pure states $|\chi\rangle $
randomly drawn (with respect to the natural Fubini -- Study measure
\cite{So01,Ha98}) from a $6M$ dimensional state space, with $M=4,6,10$.
The larger the dimension $M$, the more mixed is $\rho={\rm tr}_M(|\chi
\rangle \langle \chi |)$, hence the smaller its concurrence.
Furthermore, since ${\rm rank}(\rho)\leq M$, the $M=4$ ensemble contains
only states which do not have full rank. Fig.~\ref{ublbfig} compares the
lower bound (\ref{lowerbound}) of the concurrence of the random states
drawn from these three ensembles to its upper bound, which is obtained
as the average concurrence with respect to a particular (numerically
optimized) decomposition of the state $\rho$. Obviously, our lower bound
comes very close to the upper bound, for the entire range from strongly
entangled to almost separable states. 

In many cases, in particular for all the  relatively weakly entangled
states drawn from the $M=10$ ensemble, we have even found exact
agreement of the lower and the upper bound of the concurrence.  All
these states, which are marked by dots in Fig.~\ref{ublbfig}, fulfill
the property that there is only one entangled  substate among the three
(optimal) $2\times 2$ substates, i.e., $C(\rho^{(12)}) >0$, and
$C(\rho^{(13)}) = C(\rho^{(23)}) =0$.  Let us examine closer under which
conditions an exact agreement of upper and lower bound may then be
expected: if we reexamine (\ref{concave1}) and (\ref{concave2}) above,
with the specific choice $r^{(12)}=1$ and $r^{(13)}=r^{(23)}=0$
according to (\ref{weights}), we conclude that the lower bound in
(\ref{concave2}) can only be exact if all the states
$P^{(13)}|\psi_l\rangle$ and $P^{(23)}|\psi_l\rangle$ are product
states. If so, $P^{(12)}|\psi_l\rangle$ is either a product state, or
$|\psi_l\rangle$ lies in the image of $P^{(12)}$.  In other words, {\em
any} entangled $|\psi_l\rangle$ in an optimal decomposition lies in the
image of $P^{(12)}$.  Since for any $2\times 2$ state there exists an
optimal decomposition consisting of one entangled pure state and a
separable remainder (this follows from the derivation presented in
\cite{Wo98}), it follows that there exists an optimal decomposition of
the $2\times 3$ state $\rho$ consisting of only {\em one} entangled pure
state. This suggests the following procedure to test whether the lower
bound (\ref{lowerbound}) is saturated: if there is only one entangled
$2\times 2$ substate $\rho^{(12)}$ among the three $\rho^{(ij)}$'s, we
try to find a pure state $|\psi\rangle$ with concurrence
$C(|\psi\rangle)=C\bigl(\rho^{(12)}\bigr)$, such that
$\rho~-~|\psi\rangle\langle\psi|$ is positive and separable.  If this
succeeds, we have found the exact value of the concurrence of $\rho$.
The dots ($\bullet$) in Fig.~\ref{ublbfig}, where lower and upper bound
coincide, have been obtained in this way.

For those mixed states for which  at least two of the $2\times 2$ states
$\rho^{(ij)}$ are entangled, and equally so for some with only one
entangled projection $\rho^{(ij)}$ (in particular for the low rank
states of the $M=4$ ensemble), the lower and upper bound of the
concurrence differ, although the difference is small in most cases.
Since it is numerically quite expensive to obtain a good value of the
upper bound (and there is no guarantee that the numerically found local
minimum is also the global one), the actual value of the concurrence may
be even closer to the lower bound. 

To give an explicit example for which our lower bound can be proved to
yield the exact value of the concurrence, we use the standard product
basis, $|i,j\rangle = |i\rangle \otimes |j\rangle,\; i=1,2,\; j=1,2,3$,
to define the following family of states:
\begin{equation}
\rho_{x,y}~=~x|\psi_1\rangle\langle\psi_1|~+~y
|\psi_2\rangle\langle\psi_2|~+~\frac{1-x-y}{6}{\mathbbm 1},
\end{equation}
with
\begin{equation}
|\psi_1\rangle~=~\frac{1}{\sqrt{2}}\bigl(|11\rangle+|22\rangle\bigr),
~|\psi_2\rangle~=~\frac{1}{\sqrt{2}}\bigl(|13\rangle+|21\rangle\bigr),
~x\geq y,~x+y\leq 1.
\end{equation}
For $y=0$, these states may be considered as generalized Werner states
\cite{We89}.  By projection onto the subspace $P^{(13)}$, we obtain
\begin{equation}
C(\rho_{x,y})\geq \tilde{C}=x-\frac{1}{3}\sqrt{(1-x+2y)(1-x-y)}. 
\end{equation}
On the other hand, the state $\rho~-\tilde{C}| \psi_1\rangle \langle
\psi_1|$ is positive and separable, provided that
\begin{equation}
x\leq 1-\frac{3\sqrt{5}-1}{2}y.\label{para}
\end{equation}
Thereby, we have established $C(\rho_{x,y})=\tilde{C}$ in this parameter
regime. (If $\tilde{C}\leq 0$, which is the case for $16 x\leq
-2-y+3\sqrt{4+4y-7y^2}$, the state $\rho_{x,y}$ is separable, i.e.,
$C(\rho_{x,y})=0$.) 
 
As the simplest case which violates condition (\ref{para}), let us
finally consider $x=y=1/2$.  According to (\ref{2kto22pure}), the
concurrence of an arbitrary superposition $|\psi\rangle=\cos ( \theta
)|\psi_1\rangle+\sin (\theta )e^{i\phi} |\psi _2\rangle$ of the states
$|\psi_1\rangle$ and $|\psi_2\rangle$ reads $C^2(|\psi\rangle)= 1-\cos
^2 (\theta ) + \cos ^4 (\theta )$. The minimum $C_{\rm min} =
\sqrt{3}/2$ is realized by the two states $(|\psi _1\rangle \pm |\psi
_2\rangle)/\sqrt{2}$, which therefore form the optimal decomposition of
$\rho_{1/2,1/2}$.  Hence, we have shown that $C(\rho_{1/2,1/2}) =
\sqrt{3}/2$.  On the other hand, the lower bound (\ref{lowerbound}),
obtained by using the optimal basis $\{|1\rangle,|2\rangle,|3\rangle\}$
yields $C(\rho_{1/2,1/2}) \geq
\sqrt{\tilde{C}^2+\tilde{C}^2+0}=\sqrt{2}/2$. This proves that we cannot
expect the lower bound to be exact in all cases. We conjecture that the
above example with $x=y=1/2$ realizes the largest possible difference
between the lower bound and the actual value of the concurrence.
 
\section{Conclusion}

In summary, combining the upper bound which follows from the very
definition of mixed state concurrence with our analytical lower bound
provides a tight estimation of its exact value, for arbitrary $2\times
3$ mixed states. Furthermore, our lower bound can be efficiently
evaluated, since it involves an optimization problem on a
$K$-dimensional rather than on a $4K^2$-dimensional space. 

\acknowledgments
It is a pleasure to thank Marek Ku{\'s} and Florian Mintert for valuable
remarks and entertaining discussions, as well as R.R~Tucci for a
fruitful correspondence.  Financial support by the Volkswagen-Stiftung
and through KBN research grant (number 2P03B-072-19) are gratefully
acknowledged.

\end{document}